\documentclass[a4paper]{spie}  

\usepackage{amsmath,amsfonts,amssymb}
\usepackage{graphicx}
\usepackage[colorlinks=true, allcolors=blue]{hyperref}

\title{Multiplexable frequency retuning of MKID arrays using their non-linear kinetic inductance}

\author[a,b]{M. De Lucia}
\author[a,b]{E. Baldwin}
\author[a,b]{G. Ulbricht}
\author[a,c]{C. Bracken}
\author[b]{P. Stamenov}
\author[a]{T. P. Ray}
\affil[a]{Dublin Institute for Advanced Studies, School of Cosmic Physics, Astronomy and Astrophysics, 31 Fitzwilliam Place, D02XF86 Dublin, Ireland }
\affil[b]{School of Physics, Trinity College Dublin, College Green, Dublin 2, Ireland}
\affil[c]{Department of Experimental Physics, Maynooth University, Co Kildare, Ireland}

\authorinfo{Further author information:\\ delucia@cp.dias.ie}

\begin{document} 
\maketitle

\begin{abstract}
Microwave Kinetic Inductance Detector (MKID) arrays are currently being developed and deployed for astronomical applications in the visible and near infrared and for sub-millimetre astronomy. One of the main challenges of MKIDs is that large arrays would exhibit a  pixel yield, defined as the percentage of individually distinguishable pixels to the total number of pixels, of $75\,-\,80\,\%$ \cite{mazin2020}. Imperfections arising during the fabrication can induce an uncontrolled shift in the resonance frequency of individual resonators which end up resonating at the same frequency of a different resonator. This makes a number of pixels indistinguishable and therefore unusable for imaging. This paper proposes an approach to individually re-tune the colliding resonators in order to remove the degeneracy and increase the number of MKIDs with unique resonant frequencies. The frequency re-tuning is achieved through a DC bias of the resonator since the kinetic inductance of a superconducting thin film is current dependent and its dependence is non linear. Even though this approach has been already proposed \cite{Vissers}, our innovative pixel design may solve two issues previously described in literature such as non-negligible electromagnetic losses to the DC bias line, and the multiplexibility of multiple resonators on a single feed-line.
\end{abstract}


\section{INTRODUCTION}
\label{sec:intro}  

Microwave Kinetic Inductance Detector Arrays of up to twenty thousand pixels are deployed in instruments such as MEC\cite{Walter_2020}, DARKNESS\cite{DARKNESS} and NIKA2\cite{Calvo_2016} for astronomical applications in the visible and near infrared, and sub-millimetre bands of the electromagnetic spectrum.\\ MKIDs, being superconducting cryogenic detectors based on LC resonators, are of great interest for many astronomical applications because of their inherent spectral resolution, their single-photon sensitivity and the lack of dark counts. Astronomical instruments in the visible and near infrared equipped with MKIDs arrays can benefit from their fast time resolution, $\approx 1\,\mu s$ \cite{Day2003}, and spectral resolution, $\approx\,15$ at $400\,nm$\cite{mazin2020}, to achieve high contrasts when deployed on telescopes with Adaptive Optics Systems. At present times, the largest MKIDs array on-sky, MEC, consists of roughly $20000$ independent pixels\cite{MEC}. Proposed instruments, such as KIDSpec, will use MKIDs arrays coupled with an Echelle spectrograph with resolution $R = 4,000-10,000$ for wavelengths between $400\,nm$ and  $1500\,nm$ \cite{KIDSpec}.\\ For millimetre and sub-millimetre astronomy, MKIDs are not single-photon detecting. Given the extremely low energy band gap of the superconductors \cite{Day2003}, MKIDs can be ideal detectors for low energy applications such as Cosmic Microwave Background or millimetre astronomy and instruments such as NIKA2\cite{Calvo_2016} and DESHIMA\cite{Deshima} are incredible instruments which use Microwave Kinetic Inductance Detectors for sub-millimetre astronomy.\\
One of the challenges of MKIDs is that the fabrication yield is not optimal yet. Microscopic and macroscopic $T_C$ variations in the superconducting layer as well as inaccuracies during the fabrication process can induce significant shifts in the resonance frequency of individual resonators. Two resonators with similar resonant frequencies are practically indistinguishable and, therefore, are said to be clashing or colliding and are  considered unusable when it comes to photon detection and image reconstruction. \\
Over the last decade multiple approaches have been attempted in order to increase the yield. Among these, the $a\, posteriori$ trimming of the interdigitated fingers of the inductors so that the frequencies can be redistributed uniformly over the whole bandwidth showed great results. Another very interesting approach, as proposed in $2013$ by M.R. Vissers et al. \cite{Vissers}, exploits the non linear dependence of kinetic inductance to induce a controlled shift in resonant frequency, and therefore increasing the number of distinguishable resonators on the feed-line. 
The first approach, can, in principle, split all the resonator coupled, providing a $100\%$ yield, but requires a very difficult post-production process involving a secondary lithographic process \cite{trimming,shiboshu} or a Focused Ion Beam milling of the fingers interdigitated capacitor. Also, trimming resonators whose critical features are in the order of $1\mu $m is extremely challenging. 
The second approach, instead, can only produce a limited increment in yield, and makes the multiplexing of several pixels quite complicate. By the proposed design, all the resonators would see the effect of the biasing current and making an individual choice of resonators impossible. \\ 
In this paper, we present an improved design which would make the latter approach feasible, allowing for the selective biasing of individually chosen resonators. \\
\subsection{Kinetic Inductance Non Linearity}
\begin{figure}[h!]
    \centering
    \includegraphics[scale=0.422]{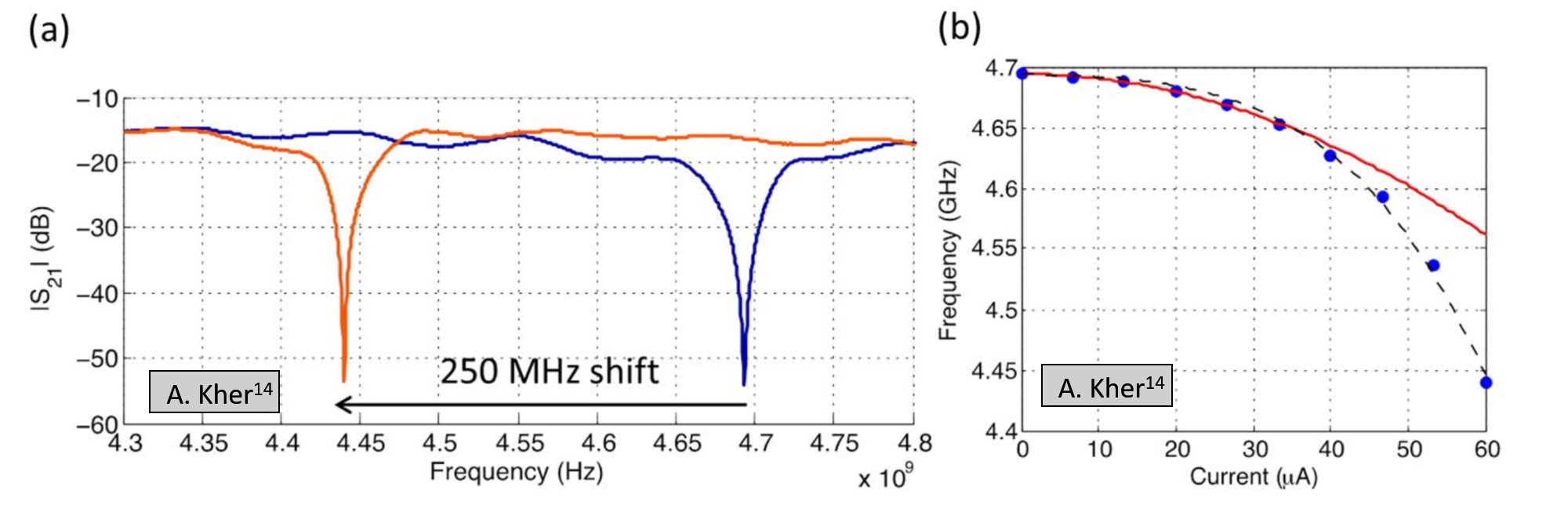}
    \caption{Reprinted from A. Kher \cite{kher}. (a): Amplitude of transmission through the chip. The blue curve shows the transmission without any current bias, the red curve shows the transmission with current bias of $60\,\mu A$. The total frequency shift is about $250\, MHz$, about $5\%$ of the resonance frequency.$\,\,$ (b) Full current response data for the  resonator. Above $60\, \mu A$, critical current of the superconductor, the resonance disappears.}
    \label{fig:kher}
\end{figure}

The electrodynamics of superconductors has been studied thoroughly and the concept of kinetic inductance is well understood, in particular Pippard\cite{Pippard1,Pippard2} described the current-dependence of the kinetic inductance which can be expanded as: 
\begin{equation}
    L_k = L_0\left(1+ \frac{I^2}{I^2_0} +...\right)
\end{equation}{}
It is clear that any odd term in the expansion has to be zero, due to symmetry reasons, and an expansion to its second, non zero term, is accurate enough to model the phenomenon. Here $I_0$ is a scaling factor and, according to literature, it is known to be of the same order of the critical current \cite{Zmuidzinas,Vissers}.\\
In the limit that the film thickness is small compared to the London penetration depth of the superconductor, the sheet inductance $L_s$ is given by:
\begin{equation}
    L_s= \frac{\hbar \rho_n}{\pi \Delta_0 t}
\end{equation}{}
Where $\rho_n$ is the normal state resistivity of the metal and $t $ is its thickness. $\Delta_0$ is the superconducting gap at $T=0\, K$ and $\pi$ and $\hbar$ are constants. A strip of length $l$ and width $w$ has a kinetic inductance of $L_k = L_s (l/w)$. The expected fractional change in kinetic inductance is given by
\begin{equation}
    \frac{\delta L_k}{L_k} =\frac{I^2}{I_0 ^2} = \kappa_*\frac{J^2}{J_0^2}
\end{equation}{}
where $\kappa_*$ is a parameter of the order of unity\cite{Zmuidzinas}, $J$ is the current density and $J_0$ 
\begin{equation}
    J_0 = \sqrt{\frac{\pi N_0 \Delta_0 ^3}{\hbar \rho_n}}
\end{equation}{}
The resonance frequency $f_r$ of an LC resonator is $1/\sqrt{LC}$, therefore a change in its nominal inductance due to a current bias, leads to a change in its resonance frequency $\delta f$. Its fractional shift, as a result of the  current bias is \cite{Vissers}.\\
\begin{equation}
    \frac{\delta f_r}{f_r} = - \frac{\delta L_{Tot}}{2L_{Tot}}= -\frac{I^2}{2I_0}
\end{equation}{}
Previous papers, \cite{Vissers,kher} showed interesting results in this direction. In both cases, a frequency shift up to $4\%$ of the designed resonance frequency was achieved, with little \cite{Vissers} to none \cite{kher} degradation in quality factor and some of these results are shown in Figure \ref{fig:kher}
{}
\section{Retuning Multiple Pixels}
\label{sec:multiple}
The two approaches described in the previous section only focus on the retuning of a single resonator. The paper by M.R. Vissers et al \cite{Vissers}. proposes a DC bias through the same feedline which is used to then read out the resonator. This approach, of course, would allow more resonators to be biased with the same DC current, but it would be impossible to selectively decide which resonators of the array to bias. Such a design would only permit a parallel shift of all the resonances towards lower frequencies. Following Adytia Kher's approach \cite{kher} which includes a DC-distribution line and allows a selective bias of the resonator, we have developed a design which incorporates all the elements just discussed and is based on a resonator geometry which is commonly known as Lumped Element Kinetic Inductance Detector or LEKID \cite{lekid}. The main advantage of such design is a wide area, described by the inductor, with high supercurrent density  which acts as the light-sensitive area and produces a rather uniform response independently on the striking position of the impinging photons. It is interesting to evaluate what improvement can be obtained when selectively biasing some of the resonators of the array. It is unimaginable that each pixel of the array, several thousand resonators per chip, is individually contacted and biased. There are two main reasons for this: The amount of cables required at the milli-kelvin stage of the cryostat would be comparable to the amount of cables needed to read out an array of Transition-Edge Sensors, therefore losing one of the main strengths of MKIDs: the frequency-domain multiplexibility of the microresonators. The second reason is that, in order to do so, the fabrication of an MKIDs array with that many biasing lines would become more and more complicated, and another strength of MKIDs would be lost.  
One acceptable compromise is to use one biasing current that is then equally redistributed between all the biased resonators; furthermore, by connecting or disconnecting individually picked resonators to the current distribution line, the fabrication yield of the array is maximised. It is crucial to investigate how big an improvement this would be and the feasibility of this approach when moving from a single pixel approach to a whole array. 

\subsection{Numerical model}
\label{sec:model}
\begin{figure}[h!]
    \centering
    \includegraphics[scale = 0.15]{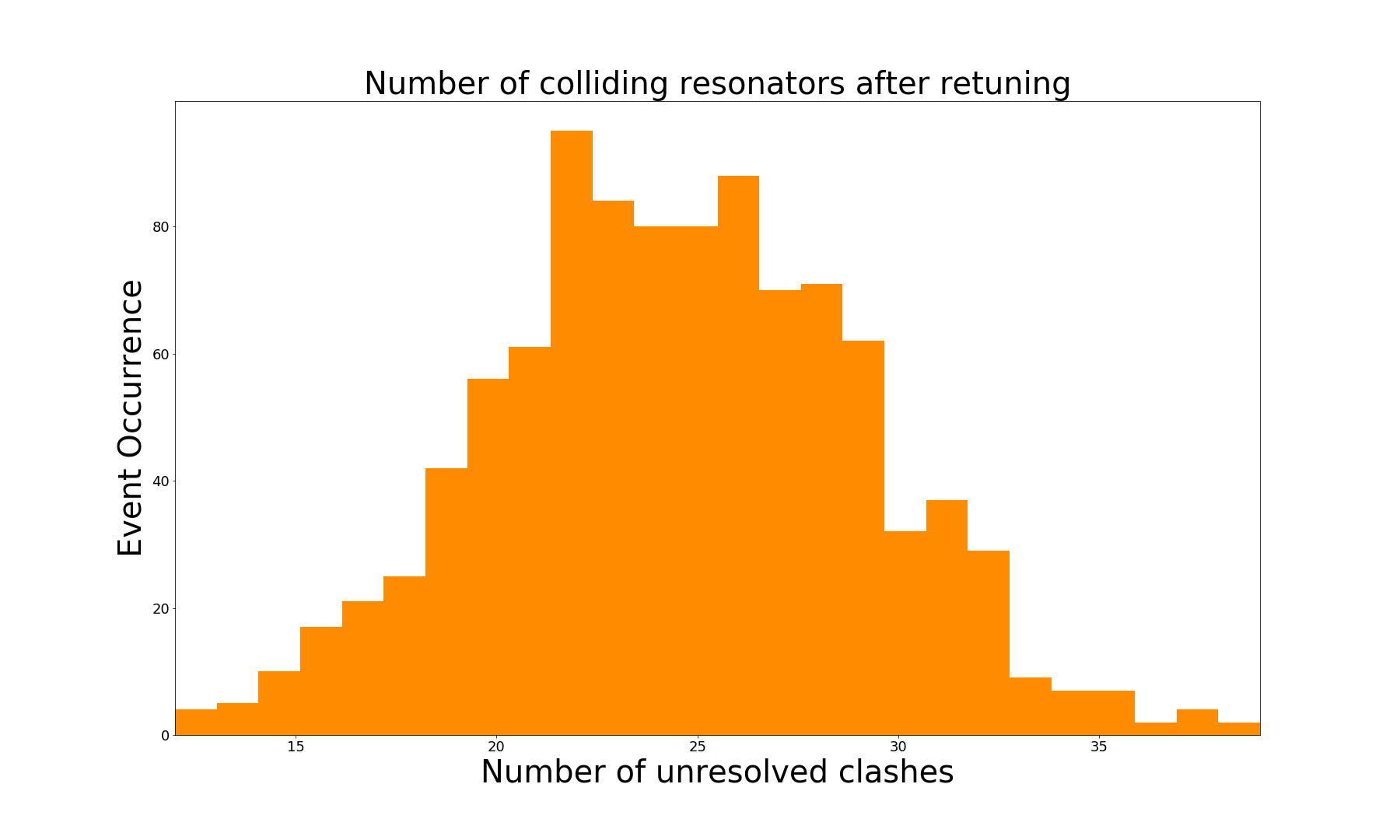}
    \caption{Number of unresolved clashes after having blindly biased one of the resonators of each pair. Starting parameters are $1000$ resonators evenly distributed over $4$ GHz bandwidth. Of the $1000$ resonators, $125$ have the same resonance frequency of $125$ others. A gaussian fit of the peak points towards a mean value of $30$ couples left to decouple}
    \label{histo_1}
\end{figure}

\begin{figure}[h!]
    \centering
    \includegraphics[scale=0.4]{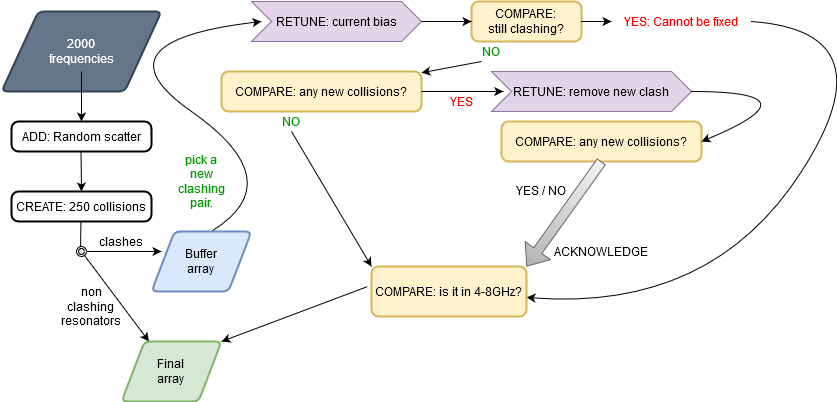}
    \caption{Flowchart describing the algorithm developed for the sake of numerical simulations }
    \label{flowchart}
\end{figure}

\begin{figure}[h!]
\begin{minipage}[b]{0.5\textwidth}
    \includegraphics[scale=0.2]{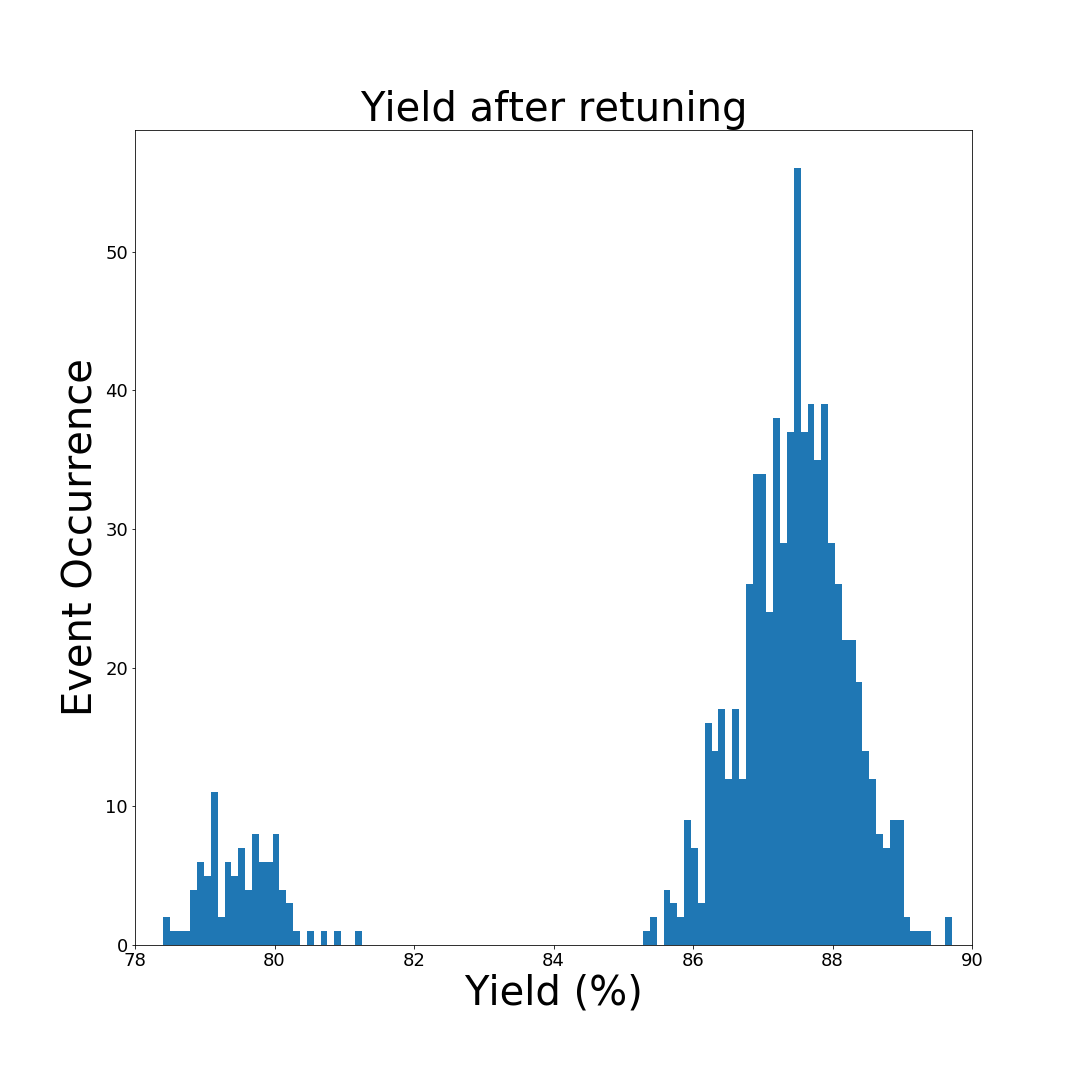}
\end{minipage}
\begin{minipage}[b]{0.5\textwidth}
  \includegraphics[scale=0.2]{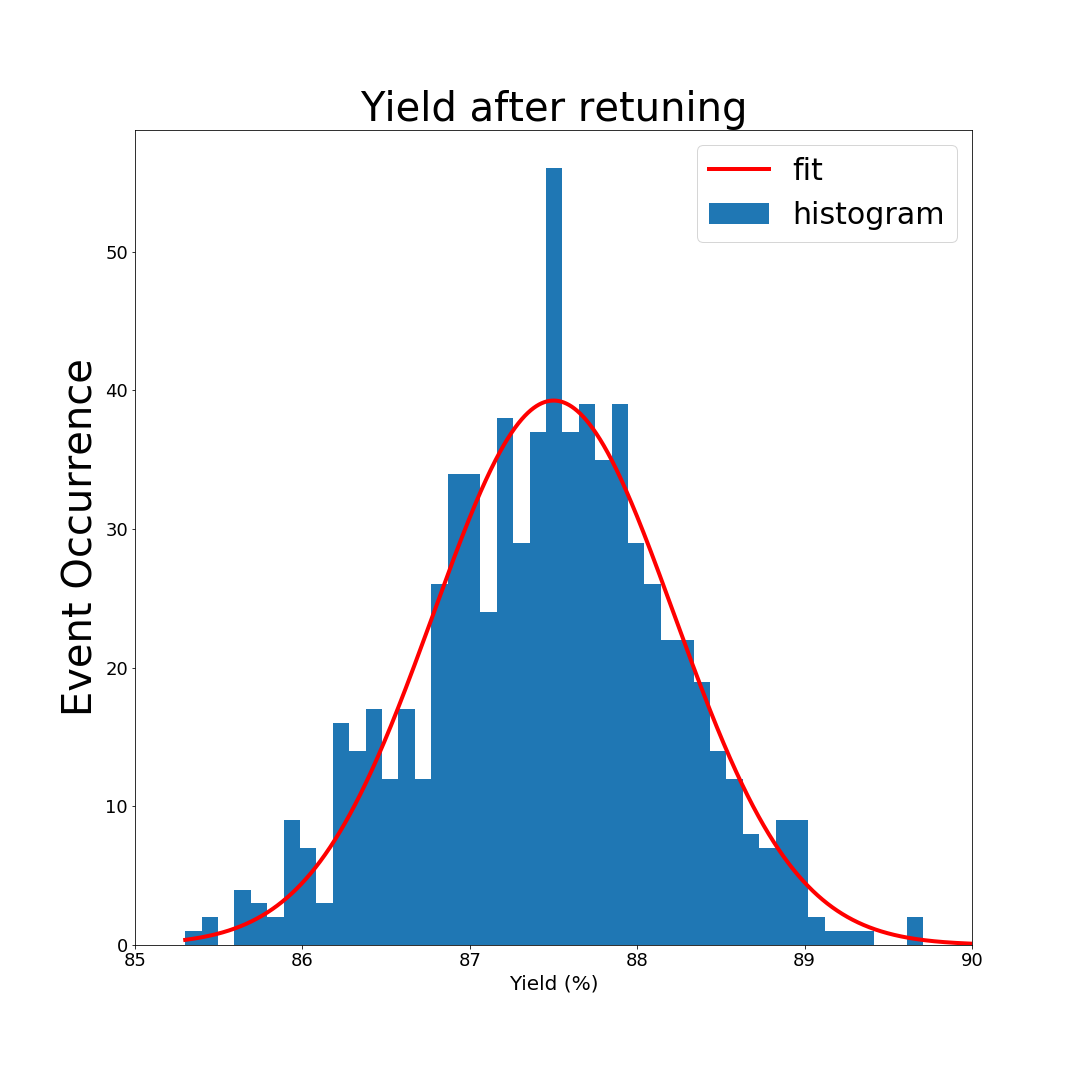}
\end{minipage}
  \caption{Left: Yield after the retuning process. A total of 800 events has been taken into account for this plot. Right: close up of the main peak from the left plot and the gaussian fit of the data.}
  \label{histo_2} 
\end{figure}

The first evaluation of the benefits of the selective retuning of individual resonators was done through a Python\cite{python} code. The code is divided in two main parts: the first one is mainly meant to verify the feasibility of the retuning by blindly biasing one of the two of clashing resonators, for every clash that occurs on the feed-line. The second part deals with the possible benefits of a smart choice of the pixels to bias.
The first part is intended to be a simple toy model in order to compare it with the second part, therefore, it only takes into account an array with $1000$ pixels per feed-line whose resonance frequencies are evenly distributed across a $4$ GHz bandwidth including 125 randomly-chosen resonators with the same exact resonance frequency as 125 other pixels. 
The programme scans through one thousand evenly spaced values of current from $10$ $\mu$A to $10$ mA and keeps track of the best case scenario, i.e. the current value by which there are the least amount of pairs of resonators which are practically indistinguishable. Usually, two resonators whose distance in frequency space is smaller than $0.5$ MHz are considered for most practical applications indistinguishable. This scan is ran one thousand times with different starting conditions to gather data with statistical significance and then a histogram is produced and it is shown in  Figure \ref{histo_1}.\\
{}
\noindent
The second stage of the process goes further to simulate a realistic MKIDs array. The number of resonators laid on each single  feed-line is now increased to current standards, $2000$ resonators evenly spaced by $2$ MHz in the $4-8$ GHz octave. therefore reducing by a factor $2$ the average spacing between two adjacent resonators compared to the first code. Furthermore, a random shift around the designed resonant frequency in the range of $\pm5$ MHz is taken into account. Only after this randomisation, any $250$ resonators are indiscriminately chosen and assigned to have the same resonant frequency of $250$ other ones. This way a $75\,\%$ starting yield is forcefully induced in the array. 
The optimisation of the retuning process is described in the block diagram in Figure \ref{flowchart} and goes as follows: \\
Starting from the colliding MKID with the highest resonant frequency, and given a value for the current bias in the range ($10\,n$A to $20\,\mu$A) five checks are performed:
\begin{enumerate}
    \item is it closer than $0.5$ MHz away from another resonator?
    \item given the bias current, what would be its new target frequency?
    \item is the collision removed?
    \item does the newly biased resonator collide with another one?  
    \item is the target frequency in the $4$-$8$ GHz octave?
\end{enumerate}
{}
The programme is meant to check the resonators in descending order, and as soon as it detects two clashing resonators, it tries to remove the degeneracy applying the current bias. Two further checks are necessary at this stage: are the two MKIDs still colliding? i.e. are the two resonators of the pair further than $0.5$ MHz away? and, \textit{in secundis}, is a new collision created with this shift? If so, the newly clashing resonator also has to be immediately re-tuned. Unfortunately, at this stage, independently whether a new collision is created, the programme, by design, will not try to remove it. This  limitation in the code is also the reason for the small peak in Figure \ref{histo_2} (Left) and a further upgrade of the code is being worked on. Now, the final check that has to be performed is making sure that all the pixels resonate within our bandwidth of interest. The upper end is of little to no interest as the process of retuning a resonator can only decrease its resonant frequency. The resonators with the lower characteristic frequencies, on the other hand, could be shifted to a different octave, which could lead to a clash between pixels when considering that second harmonic frequencies can show up in the.\\
These steps are re-iterated over different values of bias current, in order to find the current value that yields the best array possible. The whole process is re-iterated eight-hundred times in order to gather data with statistical significance. The results of this process are shown in Figure \ref{histo_2}. The left plot shows the full data set. Two peaks are clearly visible; the small peak on the right represents the events of multiple non-resolvable collisions which produce a limited improvement in overall yield. What is interesting to us, though, is the main peak on the left which as expected contains more events than the other. The plot on the right, shows a close up of said peak along with a gaussian fit of the data. \\
{}
Neglecting the smaller peak on the right, a gaussian fit of the data, estimates that the average yield of a retuned array is  $87.5\,\pm1.0\,\%$ which corresponds to a $12.5\,\%$ overall improvement in fabrication yield. A better algorithm to determine which resonators to bias and which ones to keep unbiased might further improve the overall yield by a few percents as well as a higher number of data points, possibly obtained with a current scan that is finer than $10\,n$A. An optimisation of the algorithm, which has a complexity order $O(n^3)$, is necessary and is being worked on.
\section{Resonator design}
\label{sec:design}
\begin{figure}[h!]
    \begin{minipage}[b]{0.5\textwidth}
    \centering
    \includegraphics[scale=0.6]{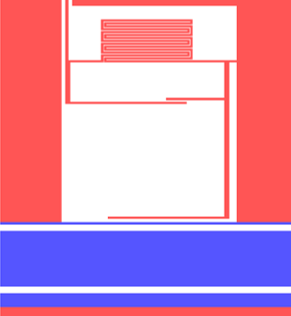}
\end{minipage}
\begin{minipage}[b]{0.5\textwidth}
  \includegraphics[scale=0.28]{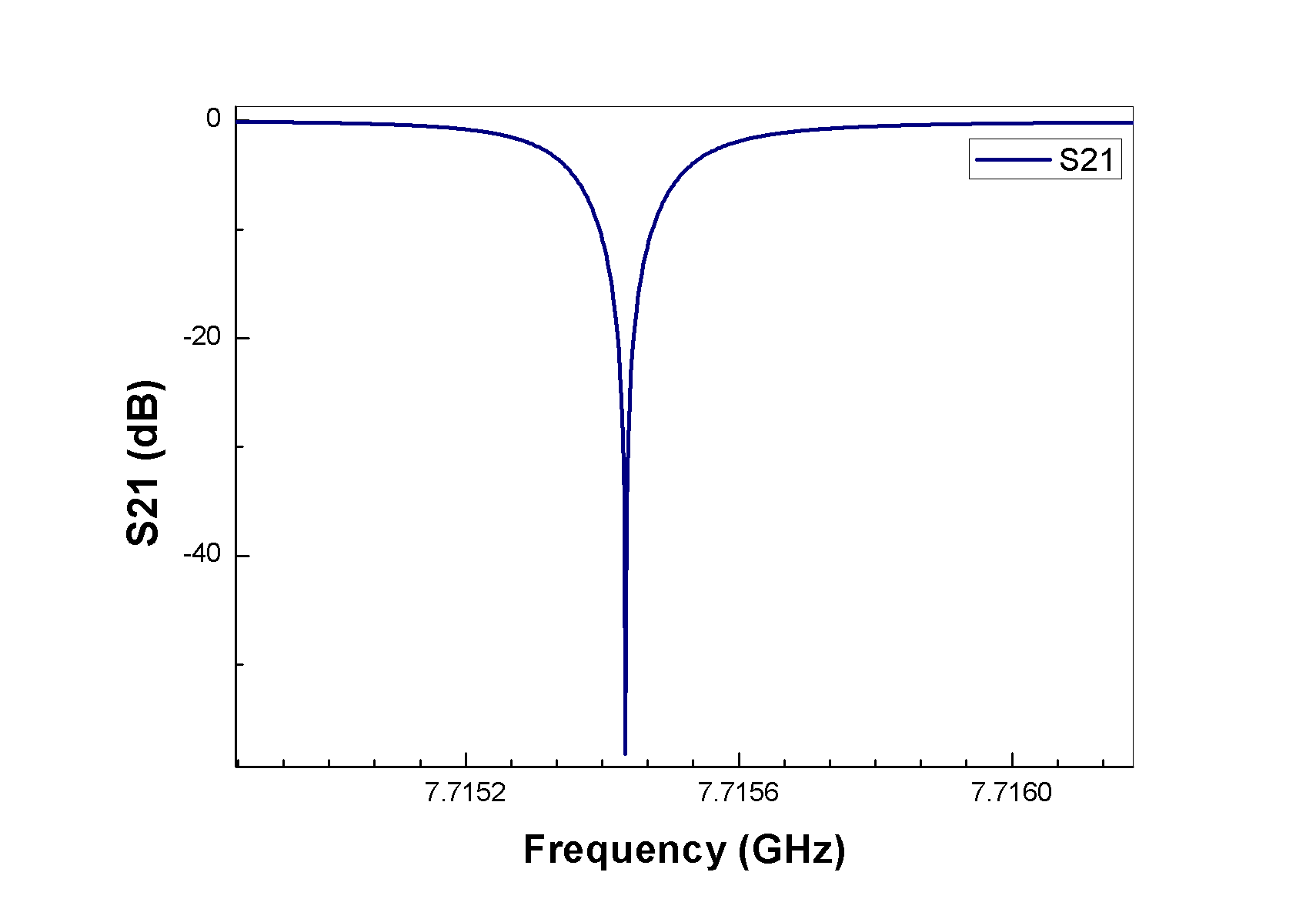}
\end{minipage}
    \caption{Left: Design of a current-biasable MKID. The standard geometry remains unchanged, a meandered inductor and an interdigitated capacitor capacitively coupled to a co-planar wave guide transmission line. In addition to that, one of the nodes of the  circuit is grounded and the other node is connected to the current distribution line through a supply line. Right: Transmission dip of the resonator ($f_0 = 7715.4$ MHz and $Q_C = 33200$)}
    \label{design_1}
\end{figure}

Although different MKID designs can be adapted so to accept a current bias, our main interest is not to give up on the \textit{Lumped Element} geometry, therefore keeping the structure of an interdigitated capacitor and a meandered inductor. In order to bias with a DC current the resonator, it can no longer rely on a floating ground at the middle point of the inductor, but a physical ground is now necessary. Furthermore, a superconducting supply line is required to connect, at will, the resonator with the current distribution line. A diagram of this geometry, drawn and analysed with the EM simulation programme SONNET\cite{sonnet} is shown in Figure \ref{design_1}. The elements previously discussed are still featured in this design, and in addition it is easy to identify the hard-wiring to the ground plane and the supply line that connects to the current distribution line. The details of this connection are intentionally not shown in this graph as further discussion is necessary. \\
Figure \ref{design_1} shows the two usual elements of the resonator: an interdigitated capacitor and a meandered inductor as well as the feed-line to which the MKID is coupled. In particular, it is interesting to notice that the feed-line and the rest of the geometry are not made of the same materials. The feed-line (in blue) is made of a material whose sheet inductance is $\approx 20\, pH/sq$ while the ground plane, the current distribution line and the resonator (in red) are designed to be made of a high kinetic inductance material, such as granular aluminium (grAl) which can have sheet inductance as high as $500\, pH/sq$ \cite{Grunhaupt}. 
{}
\subsection{Material Choice}
When imagining this work, one of the legitimate questions that can arise regards the superconducting materials that may want to be used in this project. The most important parameter taken into consideration for this crucial choice is the critical temperature of the film. Even though materials such as Hf, PtSi, $TiN_x$ have been proven to be suitable for MKIDs; since a significant current might have to flow through the superconductor, it could be beneficial to use a superconductor whose $T_C$ is higher than $1\,K$ in order to allow a higher critical current. Besides the critical temperature, a high kinetic inductance material could be of great benefit so to keep the size of the current distribution line and the pixel  as small as possible in order for them to have the smallest footprint and increase their packing capability.\\
Granular Aluminium (grAl) could be ideal for this purpose as its critical temperature ranges between $1.2$ K and $2.3$ K, and kinetic inductance values as high as $1\,nH/sq$ are easily obtained. 
\subsection{Current distribution line}
\begin{figure}[h!]
\begin{minipage}[b]{0.5\textwidth}
    \includegraphics[scale=0.3]{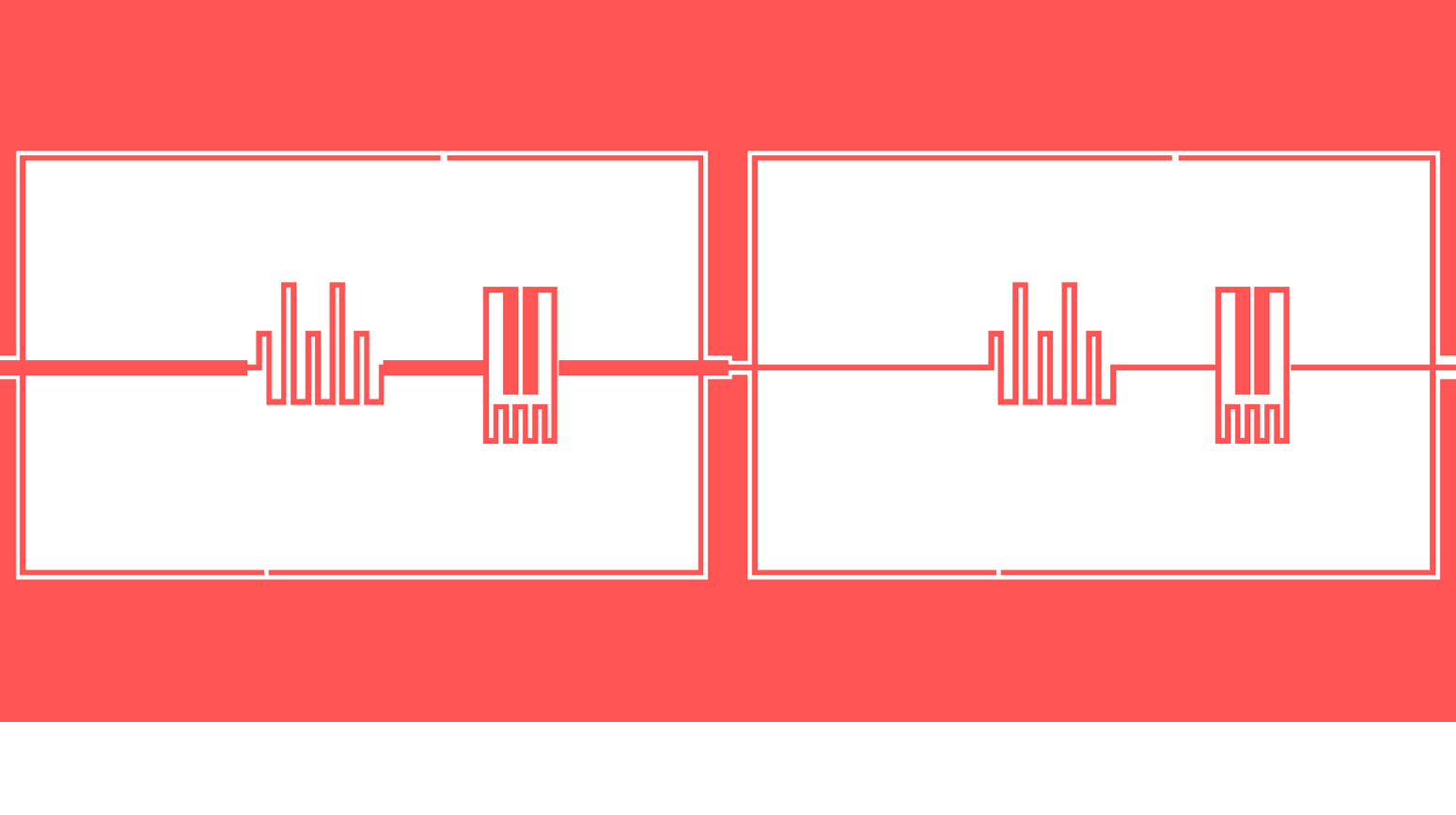}
\end{minipage}
\begin{minipage}[b]{0.5\textwidth}

  \includegraphics[scale=0.4]{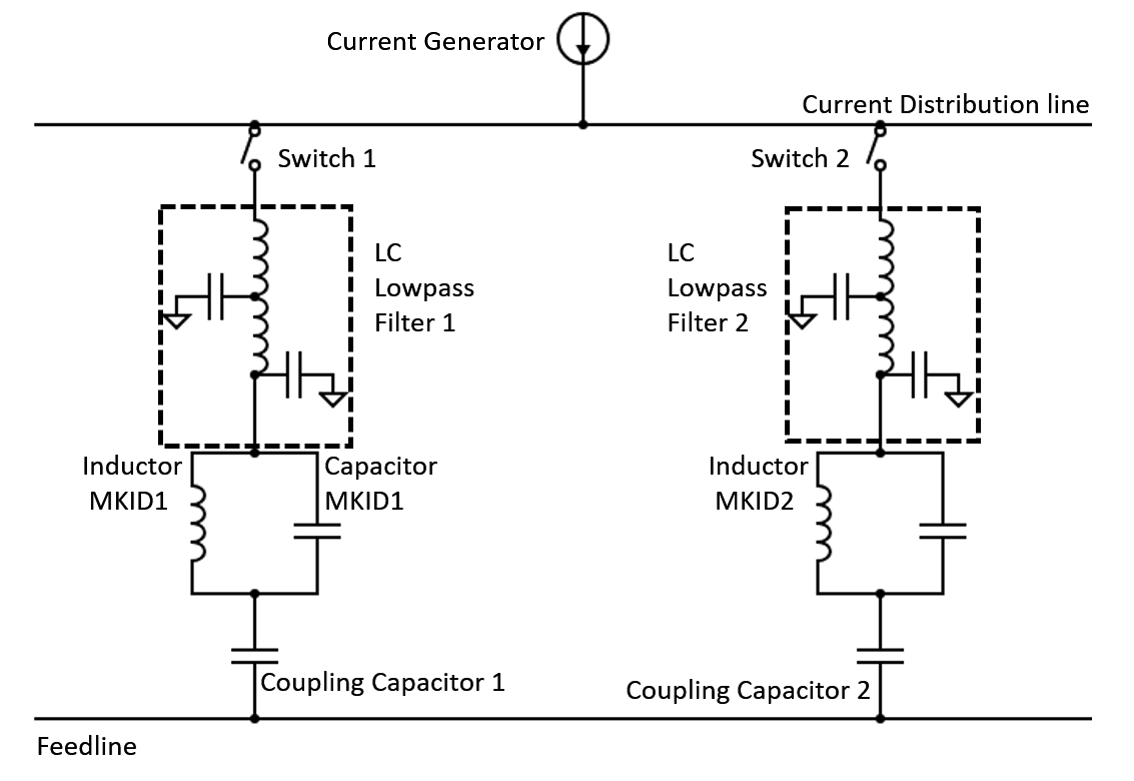}
\end{minipage}
  \caption{Left: Two stage Low pass hairpin LC filter, the cell is $300\,\mu m\times250\,\mu m$  . Right: Electrical scheme of two resonators with their current distribution line.}
  \label{design_2} 
\end{figure}

The main  function of this current distribution line is, to provide the biasing current to all the resonators in the array that need to be re-tuned . A few problems arise when implementing a current distribution line in the previously discussed geometry. \textit{In primis}, it is crucial to ensure that the high frequency energy loss is as small as possible in order to maximise the resonator's quality factor. The second, and more crucial, concern is that there is no RF signal cross-talking between any two or more pixels. In which case all the reactive elements would contribute to the creation of one single resonating circuit which would, in turn, make the whole array unusable.
In order to solve these two main issues, and following the precious advice described by A. S. Kher in \cite{kher} where the bias is achieved through a stepped-impedance filter, a multistage low-pass hairpin  LC \cite{hairpin} filter is to be considered as a viable option. Figure \ref{design_2} shows the implementation of such Low-pass filter.
{}

\begin{figure}[h!]
    \begin{minipage}[b]{0.5\textwidth}
    \centering
    \includegraphics[scale=0.3]{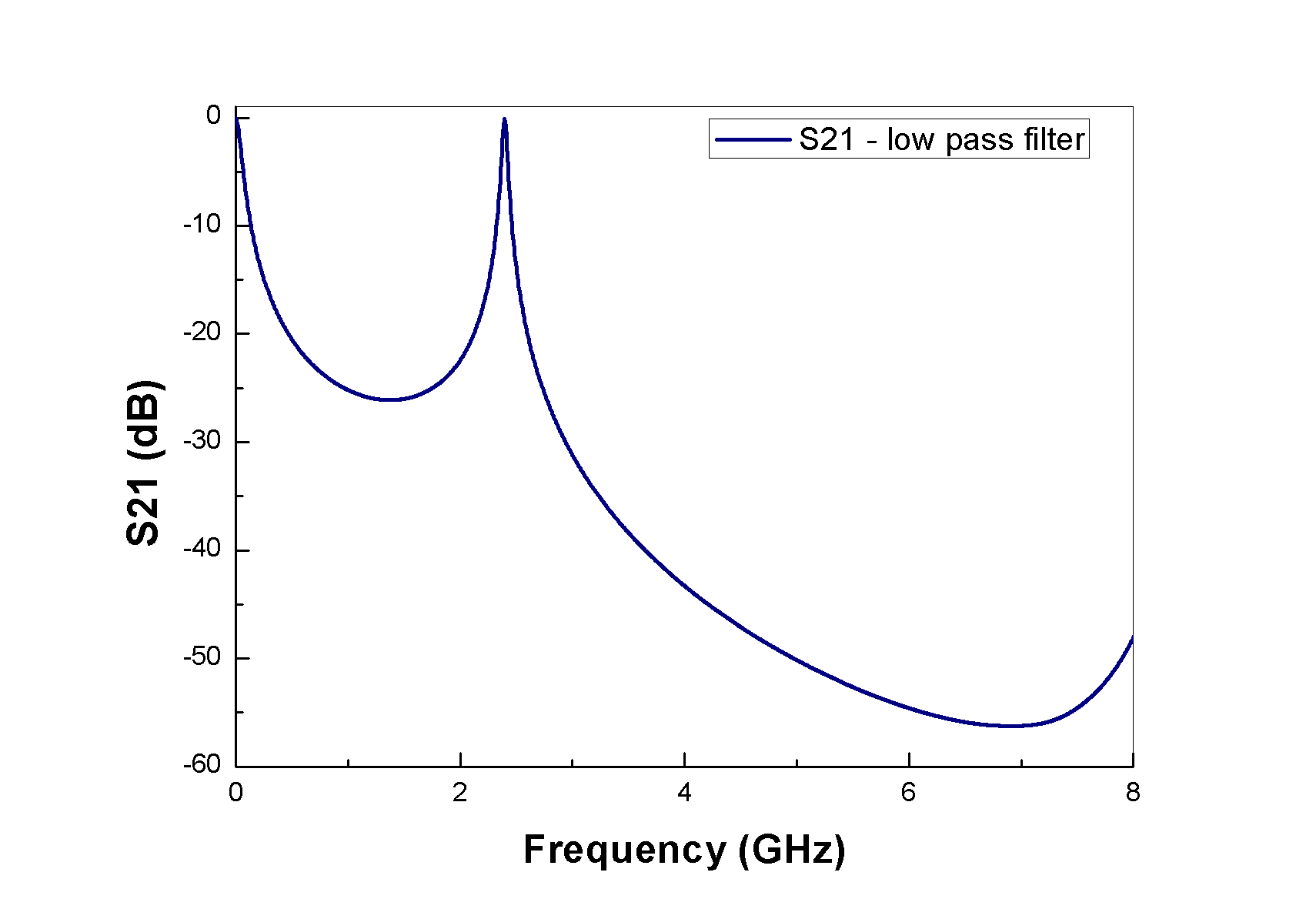}
\end{minipage}
\begin{minipage}[b]{0.5\textwidth}
  \includegraphics[scale=0.3]{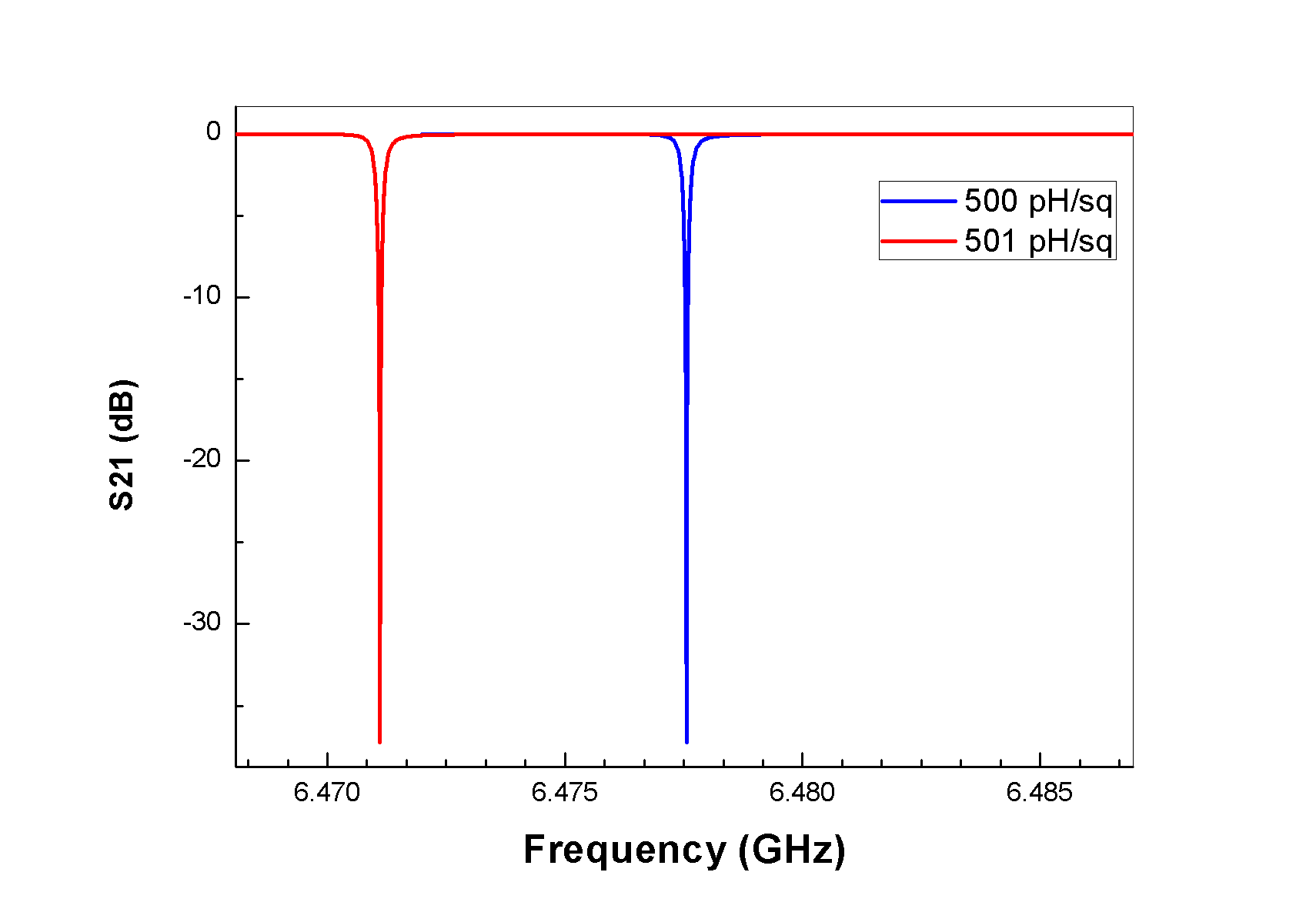}
\end{minipage}
    \caption{Left: Trans-characteristic of the two stage Low pass hairpin LC filter as simulated in SONNET. Right: SONNET simulation of  a frequency shift due to a change in Kinetic inductance. The blue curve represents the unbiased resonator ( $f_0 = 6.4775569$ GHz and $Q_C =33700$); the red curve represents the same resonator when the sheet inductance of the layer is increased by $0.2\%$ ( $f_0=6.4711046$ GHz and $Q_C=33700$).}
    \label{fig:filtering-twores1}
\end{figure}

\begin{figure}[h!]
    \begin{minipage}[b]{0.5\textwidth}
    \centering
    \includegraphics[scale=0.35]{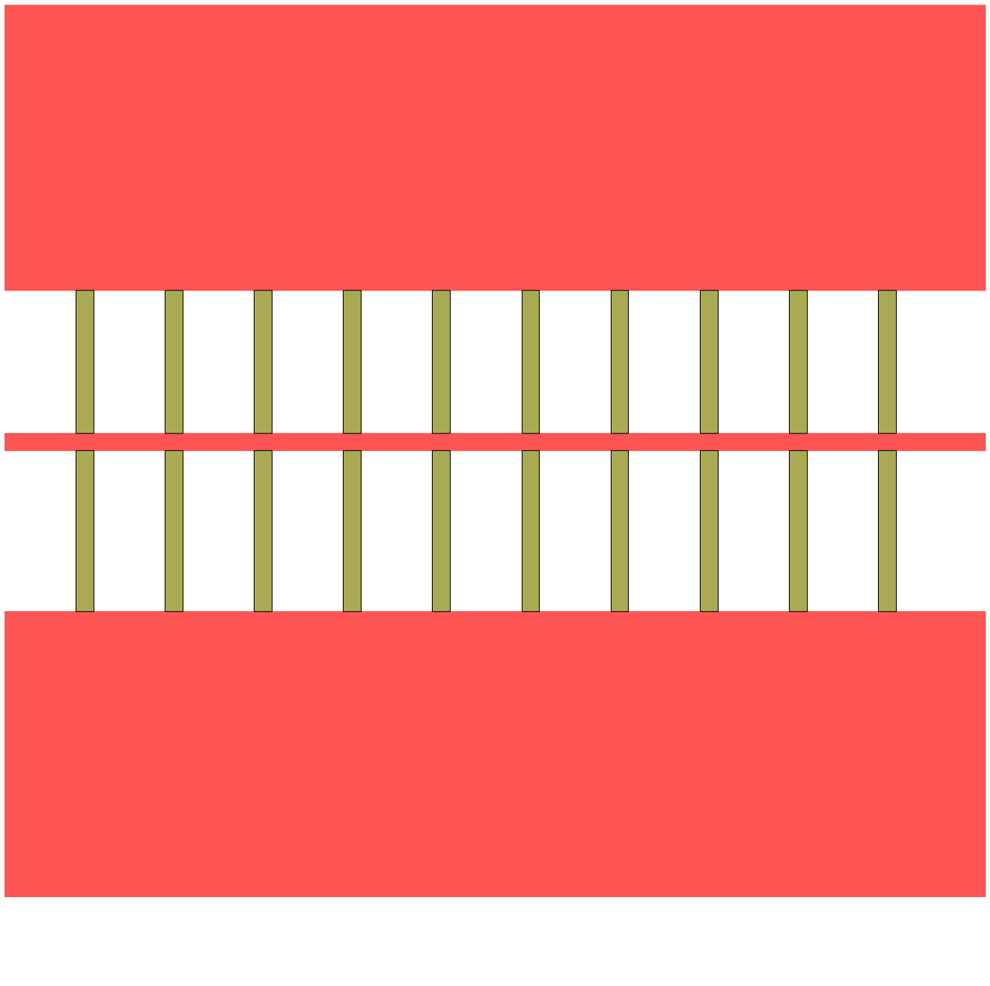}
\end{minipage}
\begin{minipage}[b]{0.5\textwidth}
  \includegraphics[scale=0.32]{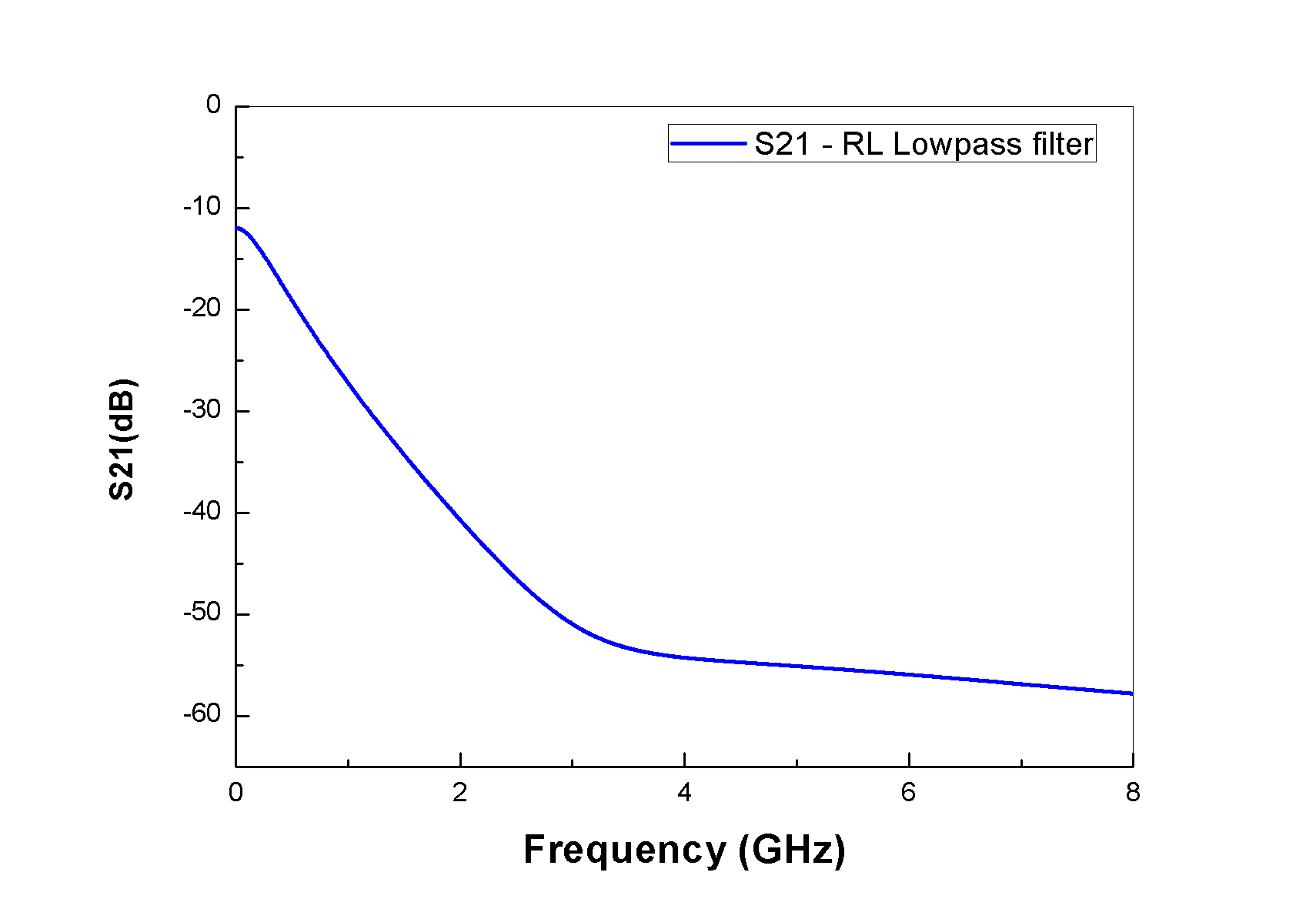}
\end{minipage}
    \caption{Left: RL low pass filter, the superconductor is depicted in red, while the yellow elements are made of gold. The cell is $50\,\mu m\times50\,\mu m$. Right: $S_{21}$ transmission of the same RL low pass filter.}
    \label{fig:filtering-twores}
\end{figure}

Given such structure, it is of great importance to know its trans-characteristics defined by its S-parameters. The main focus is to provide a filter with the lowest attenuation possible at very low frequencies, and which has at least $-40$ dB attenuation in the $4$-$8$ GHz range that is our bandwidth of operation. The S$_{21}$ curve, represented as a function of frequency, is shown in Figure \ref{fig:filtering-twores1} (Left). The attenuation is negligible at very low frequencies, the transmission rolls off  and is non significant across the $4-8$ GHz octave. For frequencies between $2.12$ GHz and $2.6$ GHz, the transmission is higher than $-20$ dB, but the presence of this peak is not relevant for the purpose of the filter. 
{}
Such implementation allows to keep two resonators decoupled while they are electrically connected through the  current distribution line. As shown in Figure \ref{fig:filtering-twores1} (Right), it is also possible to ensure a coupling quality factor which is comparable to standard MKIDs\\
Instead of showing an image of the designed MKIDs laid out on the lithographic mask as just done, when it comes to the multiplexing of different resonators on the same feedline, it is agreeable that an electrical schematic of the circuit is more efficient and a better explanatory tool. The schematic can be found in Figure \ref{design_2} (Right). \\
A hairpin LC two-stage filter is not the only possible way to achieve a wide-band low pass filter. For applications in which high packing ratios are crucial or, in case a cryostat with high cooling power is available, a RL low pass filter is also a viable option. The transmission of such filter is shown in Figure \ref{fig:filtering-twores} (Right). It is obvious that the presence of a dissipative element, here imagined as $20$ gold strips $1\mu m$ wide and $5\mu m$ long, with a sheet resistance of $20\Omega/sq$, induces a loss at low frequencies. The transmission curve shows an attenuation of $\approx12$ dB which corresponds to a four-fold attenuation of the power applied. Biasing a resonator with a current of $1\,\mu$A, which is reasonable for such applications, implies a dissipated power of $100\mu W$. Retuning a whole array of $2000$ pixels could be as energetically demanding as $25\,mW$; a cooling power only available in high performing dilution refrigerators\cite{Ahmed_2019}.
{}

\subsection{DC Bias switch}
The selection of which resonators have to be re-tuned is made by electrically connecting the resonators to the Current Distribution Line. Basically, one can physically decide whether each individual resonator has to be connected to the distribution line or not. Two approaches are here possible, one is "additive" and the other "subtractive". In the first case, the photolithographic mask would include bond pads to connect the filter to the distribution line. This way, it is possible to electrically connect the resonators  to the distribution line through well placed aluminium bond wires. Instead, the "subtractive" approach, has all the resonators already connected to the distribution line when fabricated and one can decide which resonators to disconnect. Each method has its own benefits; only having to bond the resonator one wants to re-tune is very obvious and relies on less connections, which could get damaged during the fabrication. On the contrary, the second scheme, already has all the resonators connected to the distribution line and it is possible to etch away the connections to the distribution line for those resonators that need no connection. The "subtractive" approach could be beneficial in case one wants to probe and test the array when biased with different currents. Thus it is possible to only disconnect the resonators that need be disconnected in light of  a full analysis yielding knowledge of  the behaviour of each single resonator when biased. It is important to state that, due to the inherent characteristics of the filters, when connecting or disconnecting the resonator to the  current distribution line, a shift in frequency of $ \approx 2$ MHz, in either direction, has to be taken into account, breaking the connection between the resonator and the distribution line will increase its resonant frequency without affecting the quality factor and vice-versa.
\subsection{Routing scheme}
\begin{figure}[h!]
    \centering
    \includegraphics[scale=0.12]{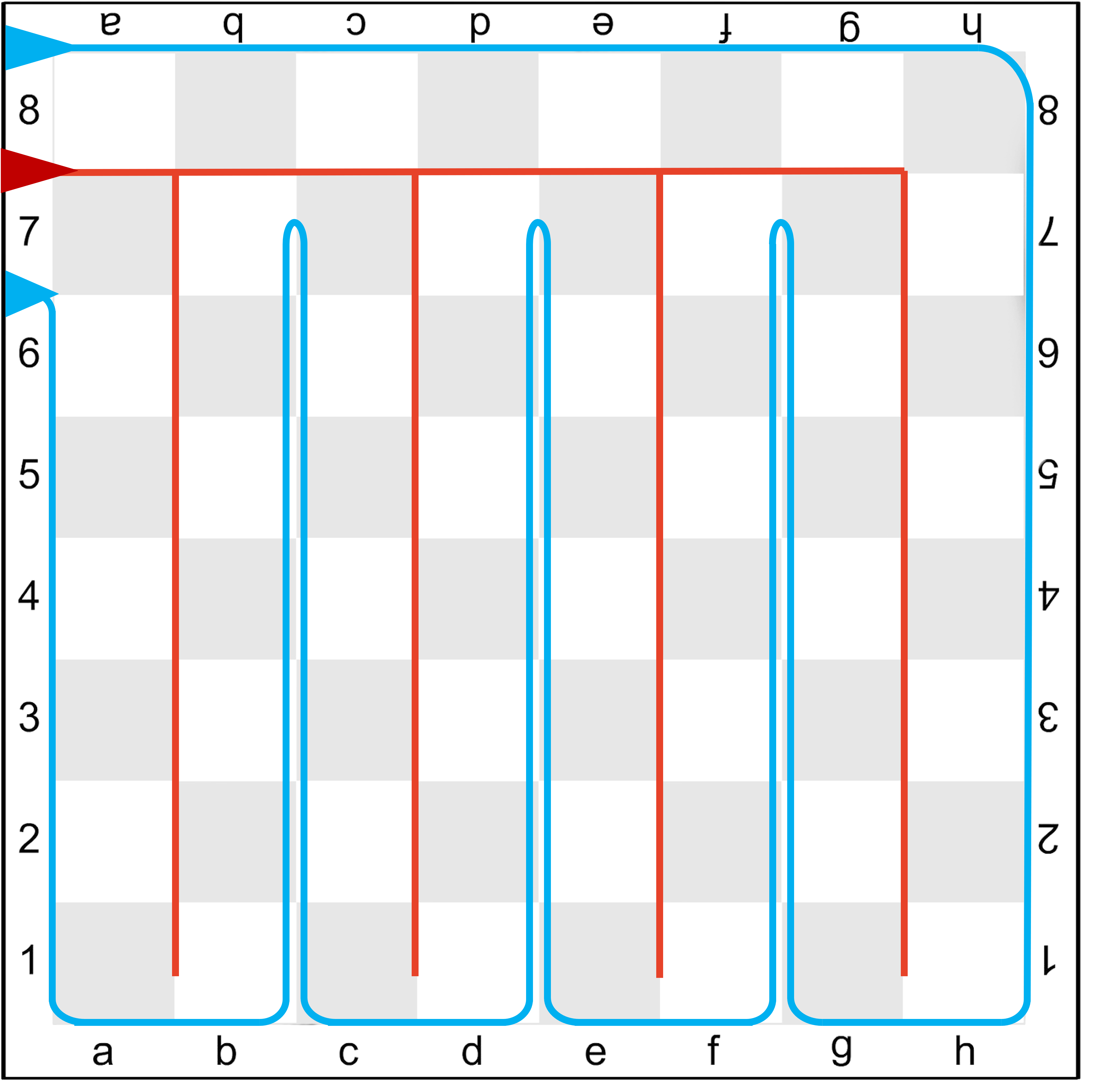}
    \caption{Routing scheme of the feed-line, in blue, and the current distribution line, in red, snaking around a generic $8    \times8$ square array. An MKID and a filter is imagined placed at the centre of each square of the chess board.  }
    \label{fig:wiring}
\end{figure}

Routing two different lines through an array of resonators in a way that the current distribution line doesn't cross the feed-line could sound difficult. Provided that a routing based on trial and error could be attempted, a Python code, based on Kruskal's algorithm \cite{cormen01introduction}, was implemented in order to find the minimum spanning tree, the shortest combination of segments that connects to all the points in a graph, that would allow the biasing of all the resonators. Then the feed-line was routed in order to be coupled to the MKIDs. At the cost of almost doubling the length of the feed-line, it is possible to route two non intersecting lines that connect to the pixels. For the sake of clarity, Figure \ref{fig:wiring} shows a $8 \times 8$ matrix with the two different routing paths. It is obvious that this wiring pattern can be extended to any rectangular array.
{}

\section{Conclusion}

Building on the state of the art of Microwave Kinetic Inductance Detectors, we proposed a possible implementation of a geometry that allows the DC bias re-tuning of an MKIDs array. Python simulations show that, with just one current value to bias the resonators with, it is possible to increase the percentage of pixels with individual frequencies from $75\%$ to above $85\%$ with minimal degradation of coupling and quality factors. The material identified as optimal for this application is granular aluminium with sheet inductance as high as $500\,pH/sq$ which is ideal in order to keep the dimensions of the filter as small as possible and, in turn, grAl is expected to withstand a higher current density when compared to superconductors with lower critical temperature. Two possible geometries have been imagined for the current distribution line, one based on a LC multi stage hairpin low pass filter, a medium-performance low dissipation filter, and the other, a single pole RL low pass filter, which has a much better performance in terms of filtering that comes with a high energetic dissipation. This makes it only suitable for high cooling power dilution refrigerators. A routing scheme, based off Kruskal minimum spanning tree, that allows the wiring of two non intersecting lines around a rectangular array of any size has been identified. Work is ongoing for first experimental tests of the devices here proposed. 
\acknowledgments 
 
This publication was emanated from research conducted with the financial support of Science Foundation Ireland under Grant number 15/IA/2880. The authors want to thank Simone Frasca for the conversation that inspired this study and  Peter Day for suggesting relevant reading material.

\bibliography{report.bib} 

\begin{thebibliography}{10}

\bibitem{mazin2020}
Mazin, B.~A., Bailey, J., Bartlett, J., Bockstiegel, C., Bumble, B., Coiffard,
  G., Currie, T., Daal, M., Davis, K., Dodkins, R., Fruitwala, N., Jovanovic,
  N., Lipartito, I., Lozi, J., Males, J., Mawet, D., Meeker, S., O'Brien, K.,
  Rich, M., Smith, J., Steiger, S., Swimmer, N., Walter, A., Zobrist, N., and
  Zmuidzinas, J., ``Optical and near-ir microwave kinetic inductance detectors
  (mkids) in the 2020s,'' (2019).

\bibitem{Vissers}
Vissers, M.~R., Gao, J., Kline, J.~S., Sandberg, M., Weides, M.~P., Wisbey,
  D.~S., and Pappas, D.~P., ``Characterization and in-situ monitoring of
  sub-stoichiometric adjustable superconducting critical temperature titanium
  nitride growth,'' {\em Thin Solid Films}~{\bf 548},  485–488 (Dec 2013).

\bibitem{Walter_2020}
Walter, A.~B., Fruitwala, N., Steiger, S., Bailey, J.~I., Zobrist, N., Swimmer,
  N., Lipartito, I., Smith, J.~P., Meeker, S.~R., Bockstiegel, C., Coiffard,
  G., Dodkins, R., Szypryt, P., Davis, K.~K., Daal, M., Bumble, B., Collura,
  G., Guyon, O., Lozi, J., Vievard, S., Jovanovic, N., Martinache, F., Currie,
  T., and Mazin, B.~A., ``The {MKID} exoplanet camera for subaru {SCExAO},''
  {\em Publications of the Astronomical Society of the Pacific}~{\bf 132},
  125005 (nov 2020).

\bibitem{DARKNESS}
Meeker, S.~R., {\em DARKNESS: The First Microwave Kinetic Inductance Detector
  Integral Field Spectrograph for Exoplanet Imaging}, PhD thesis, University of
  California, Santa Barbara (2017).

\bibitem{Calvo_2016}
Calvo, M., Benoît, A., Catalano, A., Goupy, J., Monfardini, A., Ponthieu, N.,
  Barria, E., Bres, G., Grollier, M., Garde, G., and et~al., ``The nika2
  instrument, a dual-band kilopixel kid array for millimetric astronomy,'' {\em
  Journal of Low Temperature Physics}~{\bf 184},  816–823 (Mar 2016).

\bibitem{Day2003}
Day, P.~K., LeDuc, H.~G., Mazin, B.~A., Vayonakis, A., and Zmuidzinas, J., ``{A
  broadband superconducting detector suitable for use in large arrays},'' {\em
  Nature}~{\bf 425}(6960),  817--821 (2003).

\bibitem{MEC}
Walter, A., Mazin, B.~B., Bockstiegel, C., Fruitwala, N., Szypryt, P.,
  Lipartito, I., Meeker, S., Zobrist, N., Collura, G., Coiffard, G., Strader,
  P., Guyon, O., Lozi, J., and Jovanovic, N., ``{MEC: the MKID exoplanet camera
  for high contrast astronomy at Subaru (Conference Presentation)},'' in [{\em
  Ground-based and Airborne Instrumentation for Astronomy
  VII}{\nolinebreak\hspace{0.1em}]},  Evans, C.~J., Simard, L., and Takami, H.,
  eds.,  {\bf 10702}, International Society for Optics and Photonics, SPIE
  (2018).

\bibitem{KIDSpec}
O'Brien, K., Thatte, N., and Mazin, B., ``{KIDSpec: an MKID based medium
  resolution integral field spectrograph},'' in [{\em Ground-based and Airborne
  Instrumentation for Astronomy V}{\nolinebreak\hspace{0.1em}]},  Ramsay,
  S.~K., McLean, I.~S., and Takami, H., eds.,  {\bf 9147},  143 -- 150,
  International Society for Optics and Photonics, SPIE (2014).

\bibitem{Deshima}
Hähnle, S., Marrewijk, N.~v., Endo, A., Karatsu, K., Thoen, D.~J., Murugesan,
  V., and Baselmans, J. J.~A., ``Suppression of radiation loss in high kinetic
  inductance superconducting co-planar waveguides,'' {\em Applied Physics
  Letters}~{\bf 116}(18),  182601 (2020).

\bibitem{trimming}
Liu, X., Guo, W., Wang, Y., Dai, M., Wei, L.~F., Dober, B., McKenney, C.,
  Hilton, G.~C., Hubmayr, J., Austermann, J.~E., Ullom, J.~N., Gao, J., and
  Vissers, M.~R., ``Superconducting micro-resonator arrays with ideal frequency
  spacing and extremely low frequency collision rate,'' (2017).

\bibitem{shiboshu}
Shu, S., Calvo, M., Goupy, J., Leclercq, S., Catalano, A., Bideaud, A.,
  Monfardini, A., and Driessen, E. F.~C., ``Increased multiplexing of
  superconducting microresonator arrays by post-characterization adaptation of
  the on-chip capacitors,'' {\em Applied Physics Letters}~{\bf 113}(8),  082603
  (2018).

\bibitem{Pippard1}
Pippard, A.~B. and Bragg, W.~L., ``Field variation of the superconducting
  penetration depth,'' {\em Proceedings of the Royal Society of London. Series
  A. Mathematical and Physical Sciences}~{\bf 203}(1073),  210--223 (1950).

\bibitem{Pippard2}
Pippard, A.~B. and Bragg, W.~L., ``An experimental and theoretical study of the
  relation between magnetic field and current in a superconductor,'' {\em
  Proceedings of the Royal Society of London. Series A. Mathematical and
  Physical Sciences}~{\bf 216}(1127),  547--568 (1953).

\bibitem{Zmuidzinas}
Zmuidzinas, J., ``Superconducting microresonators: Physics and applications,''
  {\em Annual Review of Condensed Matter Physics}~{\bf 3}(1),  169--214 (2012).

\bibitem{kher}
Kher, A.~S., {\em Superconducting Nonlinear Kinetic Inductance Devices}, PhD
  thesis, California Institute of Technology (2016).

\bibitem{lekid}
Doyle, S., {\em Lumped element Kinetic Inductance Detectors}, PhD thesis,
  Cardiff University (2008).

\bibitem{python}
Van~Rossum, G. and Drake~Jr, F.~L.,  [{\em Python
  tutorial}{\nolinebreak\hspace{0.1em}]}, Centrum voor Wiskunde en Informatica
  Amsterdam, The Netherlands (1995).

\bibitem{sonnet}
``Sonnet.'' \url{www.sonnetsoftware.com}.

\bibitem{Grunhaupt}
Grünhaupt, L., Maleeva, N., Skacel, S.~T., Calvo, M., Levy-Bertrand, F.,
  Ustinov, A.~V., Rotzinger, H., Monfardini, A., Catelani, G., and Pop, I.~M.,
  ``Loss mechanisms and quasiparticle dynamics in superconducting microwave
  resonators made of thin-film granular aluminum,'' {\em Physical Review
  Letters}~{\bf 121} (Sep 2018).

\bibitem{hairpin}
Cho, J.-H. and Lee, J.-C., ``Compact microstrip stepped-impedance hairpin
  resonator low-pass filter with aperture,'' {\em Microwave and Optical
  Technology Letters}~{\bf 46}(6),  517--520 (2005).

\bibitem{Ahmed_2019}
Ahmed, M., Alarcon, R., Aleksandrova, A., Bae{\ss}ler, S., Barron-Palos, L.,
  Bartoszek, L., Beck, D., Behzadipour, M., Berkutov, I., Bessuille, J.,
  Blatnik, M., Broering, M., Broussard, L., Busch, M., Carr, R., Cianciolo, V.,
  Clayton, S., Cooper, M., Crawford, C., Currie, S., Daurer, C., Dipert, R.,
  Dow, K., Dutta, D., Efremenko, Y., Erickson, C., Filippone, B., Fomin, N.,
  Gao, H., Golub, R., Gould, C., Greene, G., Haase, D., Hasell, D., Hawari, A.,
  Hayden, M., Holley, A., Holt, R., Huffman, P., Ihloff, E., Imam, S., Ito, T.,
  Karcz, M., Kelsey, J., Kendellen, D., Kim, Y., Korobkina, E., Korsch, W.,
  Lamoreaux, S., Leggett, E., Leung, K., Lipman, A., Liu, C., Long, J.,
  MacDonald, S., Makela, M., Matlashov, A., Maxwell, J., Mendenhall, M., Meyer,
  H., Milner, R., Mueller, P., Nouri, N., O{\textquotesingle}Shaughnessy, C.,
  Osthelder, C., Peng, J., Penttila, S., Phan, N., Plaster, B., Ramsey, J.,
  Rao, T., Redwine, R., Reid, A., Saftah, A., Seidel, G., Silvera, I., Slutsky,
  S., Smith, E., Snow, W., Sondheim, W., Sosothikul, S., Stanislaus, T., Sun,
  X., Swank, C., Tang, Z., Dinani, R.~T., Tsentalovich, E., Vidal, C., Wei, W.,
  White, C., Williamson, S., Yang, L., Yao, W., and Young, A., ``A new
  cryogenic apparatus to search for the neutron electric dipole moment,'' {\em
  Journal of Instrumentation}~{\bf 14},  P11017--P11017 (nov 2019).

\bibitem{cormen01introduction}
Cormen, T.~H., Leiserson, C.~E., Rivest, R.~L., and Stein, C.,  [{\em
  Introduction to Algorithms}{\nolinebreak\hspace{0.1em}]}, The MIT Press,
  2nd~ed. (2001).

\end{thebibliography}
\bibliographystyle{spiebib} 

\end{document}